# COS OBSERVATIONS OF METAL LINE AND BROAD LYMAN-α ABSORPTION IN THE MULTI-PHASE O VI and Ne VIII SYSTEM AT z = 0.20701 TOWARD HE 0226-4110[1]

B. D. Savage[2], N. Lehner[3], and A. Narayanan[4]

## ABSTRACT

Observations of the QSO HE 0226-4110 ($z_{em}$ = 0.495) with the Cosmic Origins Spectrograph (COS) from 1134 to 1796 Å with a resolution of ~17 km s$^{-1}$ and signal-to-noise (S/N) per resolution element of 20 to 40 are used to study the multi-phase absorption system at z = 0.20701 containing O VI and Ne VIII. The system was previously studied with lower S/N observations with FUSE and STIS. The COS observations provide more reliable measures of the H I and metal lines present in the system and reveal the clear presence of broad Lyman α (BLA) absorption with b = 72(+13, -6) km s$^{-1}$ and logN(H I) = 13.87±0.08. Detecting BLAs associated with warm gas absorbers is crucial for determining the temperature, metallicity and total baryonic content of the absorbers. The BLA is probably recording the trace amount of thermally broadened H I in the collisionally ionized plasma with log T ~5.7 that also produces the O VI and Ne VIII absorption. The total hydrogen column in the collisionally ionized gas, logN(H) ~ 20.1, exceeds that in the cooler photoionized gas in the system by a factor of ~22. The oxygen abundance in the collisionally ionized gas is [O/H] = -0.89±0.08±0.07. The absorber probably occurs in the circumgalactic environment (halo) of a foreground L = 0.25L$_*$ disk galaxy with an impact parameter of 109h$_{70}^{-1}$ kpc identified by Mulchaey & Chen (2009).

*Short title: The O VI/Ne VIII System at z = 0.20701 Toward HE 0226-4110*

[1] Based on observations obtained with the NASA/ESA Hubble Space Telescope, which is operated by the Association of Universities for Research in Astronomy, Inc. under NASA contract NAS5-26555, and the NASA-CNES/ESA Far Ultraviolet Spectroscopic Explorer mission operated by Johns Hopkins University, supported by NASA contract NAS 05-32985.

[2] Department of Astronomy, University of Wisconsin-Madison, 475 North Charter Street, Madison, Wi 53706
[3] Department of Physics, University of Notre Dame, 225 Nieuwland Science Hall, Notre Dame, IN 46556
[4] Indian Institute of Space Science and Technology, Thiruvananthapuram 695547, Kerala, India

1. INTRODUCTION

Determining the physical conditions and abundances in intervening QSO absorption line systems containing highly ionized atoms such as O VI and Ne VIII has been hampered by the difficulty of detecting broad but weak H I absorption associated with trace amounts of H I in the highly ionized gas in those systems with log T(K) > 5. Without reliable information on the physical and ionization conditions in the absorbers, it is difficult to use the absorbers to draw firm conclusions about their baryonic content. The broad Lyman α (BLA) absorption features in highly ionized QSO absorption line systems are weak and often confused with absorption produced in cooler photoionized gas in the same absorption system. The direct evidence that some of the O VI absorption systems are tracing gas with log T > 5 rather than photoionized gas with log T ~ 4.3 is beginning to be obtained through the careful study of a limited number of O VI absorption systems (see Table 1 in Savage et al. 2011). In contrast, it has been easier to show that ~25% of the O VI absorption components observed at low redshift likely arise in low density photoionized gas with log T ~ 4.3. This is because the associated H I absorption in the photoionized systems is strong, relatively narrow with b(H I) < 30 km s$^{-1}$, and therefore relatively easy to detect[5] (Tripp et al. 2008; Thom & Chen 2008).

In contrast for gas with 5 < logT < 6, the associated BLAs will have Doppler parameters, b(H I), ranging from 40 to 130 km s$^{-1}$ and hydrogen column densities, logN(H I) ranging up to ~14. Relatively high S/N spectra are required to convincingly detect the BLAs. In several cases, the presence of Ne VIII in absorption systems containing O VI requires an origin in collisionally ionized gas with log T > 5 because Ne VIII is very difficult to produce in photoionized gas (Savage et al. 2005; Narayanan et al. 2009 ; Narayanan et al. 2010). While the detection of Ne VIII is important and points to the presence of gas with log T ~ 5.5-6.0, it is necessary to assume a value for the elemental abundance of Ne and/or O in the system in order to convert the measured column densities of Ne VIII and O VI into measures of the total associated column density of H = H$^o$ + H$^+$. This same abundance assumption must also be made by X-ray astronomers attempting to measure the baryonic content of the warm-hot intergalactic medium (WHIM) through X-ray absorption line spectroscopy of O VII and O VIII. However, the abundances of the metals in these low redshift absorbers could range from well less than 0.1 solar to several times solar. With the common assumption that the abundance is ~0.1 solar, the uncertainty in the baryon estimate could be as large as a factor of ~10. Therefore, measures of the BLAs associated with high ionization metal absorption lines systems are essential for ultimately determining the baryonic content of these systems.

If the BLA associated with O VI (and/or Ne VIII) absorption systems can be detected, measures of the different values of the Doppler parameter for the H I and the heavier O VI or NeVIII can be used to estimate the temperature of the gas. The gas temperature and an understanding of the ionization properties of the gas then allows measures of logN(H I) for the BLA to be converted into measures of the total column of hydrogen logN(H) in the absorption system. Such an estimate has been obtained for the strong O VI system toward HE 0153-4520 at z = 0.22601 detected by Savage et al. (2011). In that system a BLA with b(H I) = 140(+14, -16) km s$^{-1}$ and logN(H I) = 13.70(+0.05, -0.08) is likely associated with O VI having b = 37±1 km s$^{-1}$ and logN(O VI) = 14.21±0.02. The O VI and BLA observations imply the direct detection of thermally broadened H I in hot gas with log T = 6.07(+0.09, -0.12), [O/H] = -0.28(+0.09, -0.08) and logN(H) = 20.41(+0.13, -0.17). The baryonic column density in the hot[6] plasma traced by this absorption system is very large.

---

[5] We note that the existence of aligned O VI and narrow H I absorption does not guarantee the O VI arises in the narrow H I component. If the S/N at Lyman α is relatively low, a BLA could be superposed on the narrow H I absorption and remain undetected. In fact narrow and broad H I absorption features have been detected in COS spectra (see Savage et al. 2011).

The detection of the very broad BLA in the system toward HE 0153-4520 has been made possible by the recent deployment of the Cosmic Origins Spectrograph (COS) on the Hubble Space Telescope (HST). COS provides a major new facility for studies of the low redshift IGM. Although COS has a spectral resolution several times lower than the E140M mode on the Space Telescope Imaging Spectrograph (STIS), COS is 10-20 times more efficient in collecting photons than STIS, making it possible to obtain relatively high S/N spectra of bright QSOs for absorption line studies of the IGM. With high S/N observations, the search for broad H I absorption associated with O VI and Ne VIII absorption is greatly facilitated.

In this paper we report on COS observations of the z = 0.20701 system in the spectrum of the bright QSO HE 0226-4110 containing O VI and Ne VIII tracing gas estimated to have log T ~ 5.7 along with many other metal lines tracing cooler photoionized gas. The properties of the gas in this system have been previously studied by Savage et al. (2005) using observations from FUSE and STIS (see section 3). The absorber probes the circumgalactic environment of a galaxy system containing two galaxies with L = 0.25L$_*$ and one with L = 0.05L$_*$, all within Δv < 200 km s$^{-1}$ and ρ < 290h$_{70}^{-1}$ physical kpc of the absorber (Mulchaey & Chen 2009). The properties of these galaxies and their relationships to the absorber are discussed in section 8. The new COS observations with S/N per resolution element ranging from 20 to 40 over the 1134 to 1794 Å wavelength range have much higher S/N than the STIS observations of Savage et al. (2005). The new observations allow improved measurements of the properties of the metal lines in the system and reveal the presence of a BLA associated with the O VI and Ne VIII in the system with b(H I) = 73 (+12, -5) km s$^{-1}$ and logN(H I) = 13.87±0.08. The detection of the BLA allows for an improved understanding of the ionization conditions in the absorber and permits an estimate of the metallicity and total baryonic content of the warm plasma in the system.

Distances in this paper are physical distances assuming a ΛCDM cosmology with Ω$_M$ = 0.3, Λ = 0.7 and H$_0$ = 70 km s$^{-1}$ Mpc$^{-1}$ with h$_{70}$ = H$_0$/ 70 km s$^{-1}$.

## 2. COS OBSERVATIONS

HE 0226-4110 was observed by COS in February of 2010 (see Table 1). We have combined the COS G130M and G160M observations spanning the wavelength range from 1137 to 1796 Å. Information about COS can be found in Froning & Green (2009), and the COS HST Instrument Handbook (Dixon et al. 2010). The inflight performance of COS is discussed in Osterman et al. (2011) and in numerous instrument science reports found on the Space Telescope Science Institute COS website at
http://www.stsci.edu/hst/cos/documents/isrs.

---

[6] In this paper we adopt the temperature naming convention for IGM research where cool, warm, and hot refer to plasmas with log T ~ 4 to 4.7, 5 to 6, and > 6, respectively. The cool plasmas are likely photoionized while the warm and hot plasmas are mostly collisionally ionized.

The spectral integrations were obtained with different grating set up wavelengths in order to obtain spectra with different detector/wavelength alignments to reduce the effects of detector fixed pattern noise and provide a way of covering the wavelength gaps between the two detector micro-channel plate segments (A and B). The various integrations and their HST MAST identification code, dates of observation, and integration times are given in Table 1.

The micro-channel plate delay line detector was operated in the time-tag mode with the QSO centered in the 2.5" diameter primary science aperture. The internal wavelength calibration lamps were flashed several times during each science integration. The time-tag data were processed with CalCOS with the reference files available 24 February 2010. A comparison of this extraction to those using more recent versions of CalCOS reveals no significant changes to the quality of the extracted spectra except for modifications to the wavelength calibrations. However, the wavelength calibration changes do not affect our program because the COS wavelengths for the HE226-4110 observations are adjusted to those provided by the STIS observations.

The scattered light background levels in COS spectra are very small and unimportant. Flat-fielding, alignment and co-addition of the individual processed spectra utilized software developed by the COS team for the processing of FUV observations as described by Danforth et al. (2010).

The effects of major detector defects were removed when producing the combined spectrum by giving the affected wavelength regions in each individual spectrum low weight during the addition process. The reduced intensity in the grid wire shadows was also corrected for in the individual integrations and these affected wavelength regions were given lower weight in the co-addition process.

The proper alignment of the individual spectra was achieved through a cross-correlation technique. The different individual spectra (in flux units) were weighted by integration time when combined.

The S/N per ~17 km s$^{-1}$ resolution element for the resulting combined spectrum spans the range from 20 to 40 for 1150< λ < 1750 Å with the peak S/N occurring in the 1400 to 1500 Å range. Since the weaker fixed pattern noise features were not flat fielded but simply averaged, there are residual features at the 2-3% level throughout the spectrum. The S/N was directly measured in the observed combined spectra from the rms error in various regions of the continuum. The measured S/N therefore includes contributions from the fixed pattern noise.

A characterization of the COS line spread function (LSF) is found in Ghavamian et al. (2009) and updated by Kriss et al. (2011). The LSF has a narrow core and broad wings. At 1200, 1300, and 1400 Å the LSF has a full width at half maximum (FWHM) in velocity of 17.1, 15.4, and 13.9 km s$^{-1}$, respectively. However, the broad wings on the LSF have FWHM ~50 km s$^{-1}$ and contain ~20 to 30% of the LSF area. The broad wing contribution is largest at the shortest wavelengths. We adopt the Kriss et al. (2011) LSF for the profile fit analysis discussed in § 4.

The COS G130M and G160M pixel sampling widths of 0.010 to 0.012 Å, provides ~ 6 to 8 samples per FWHM of the spread function. Wavelengths, velocities and redshifts reported in this paper are heliocentric. The COS wavelength calibration is obtained through Pt lamp exposures obtained as part of each integration. We used the standard COS pipeline processing for the intitial wavelength calibration of the COS spectra. However, since we also have STIS E140M observations of HE 0226-4110 from Savage et al. (2005) and Lehner et al. (2006) with superior wavelength calibration, we adjusted the COS observations to agree with the STIS observations when deriving absorption redshifts and velocities from the observations and when performing Voigt profile fits. The STIS wavelength calibration is excellent with a zero point uncertainty of ~1 km s$^{-1}$ and occasional errors of ±3 km s$^{-1}$. We estimate the velocity uncertainties in the COS observations adjusted into the STIS heliocentric reference frame to be ~±5 km s$^{-1}$. In wavelength regions where there are no clearly detected features in the STIS spectra the velocity uncertainties are larger, ~ ±10 km s$^{-1}$.

## 3. RESULTS FROM THE FUSE AND STIS OBSERVATIONS

The z = 0.20701 multi-phase absorber in the spectrum of HE 0226-4110 was first studied by Savage et al. (2005) using observations from FUSE and STIS at 20 and 8 km s$^{-1}$ resolution, respectively. Velocity plots and profile fits of the absorption features observed by FUSE and STIS are displayed in Figures 2 and 4 of Savage et al. (2005). The multi-phase system is of great interest because of the presence of strong O VI $\lambda\lambda$1031,1037 absorption with logN(O VI) = 14.37±0.03 along with a clear 3.9$\sigma$ detection of absorption by the Ne VIII $\lambda\lambda$770,780 doublet with logN(Ne VIII) = 13.89±0.11. The O VI and Ne VIII are likely produced by collisional ionization in a plasma with logT ~ 5.7.

The moderately ionized gas in the system traced by H I $\lambda\lambda$1215 to 923, C III $\lambda$977, O III $\lambda$833, N III $\lambda$989 Si III $\lambda$1206, O IV $\lambda$787 and S VI $\lambda$933 probably occurs in a gas phase photoionized by the extragalactic background radiation with a solar abundance radio of heavy elements and an overall metallicity of [Z/H] = -0.5±0.2. Simple photoionization modeling by Savage et al. implied the moderately ionized gas has T ~2.1x10$^4$ K, $n_H$ ~ 2.6x10$^{-5}$ cm$^{-3}$, P/k ~ 0.5 cm$^{-3}$, N(H) ~ 4.6x10$^{18}$ cm$^{-2}$, and N(H I)/N(H) ~ 2.5x10$^{-4}$.

We draw on these results when discussing the additional information provided by the new COS observations in the subsequent sections.

## 4. RESULTS FROM THE COS OBSERVATIONS

In Figure1 COS and STIS observations of HE 0226-4110 over the wavelength range from 1235 to 1255 Å are displayed showing H I $\lambda$1025 and O VI $\lambda\lambda$1031, 1037 absorption in the z = 0.20701 system. The other strong lines from the ISM and other IGM systems are identified in Savage et al. (2005) and Lehner et al. (2006). The large improvement in S/N from the STIS to the COS observations permits a more accurate determination of the properties of the gas in the absorption system than was possible with the FUSE and STIS measurements.

Continuum normalized velocity plots of selected absorption lines from the COS and STIS observations of the z = 0.20701 system are shown in Figure 2. While the COS observations have lower resolution, the increased S/N is important for clearly detecting the weaker absorbers such as N V λλ1238, 1242, Si III λ1206, N III λ989 and for more reliably determining the structure of the stronger absorption lines particularly H I λλ1215, 1025 and the O VI λλ1031, 1037 doublet. In the continuum near the redshifted O VI λ1031 line S/N = 11 for the STIS observations in 6.5 km s$^{-1}$ velocity bins compared to S/N = 36 for the COS observations in 15 km s$^{-1}$ velocity bins. The values in the continuum for the H I λ1215 line are S/N = 12 for STIS for 6.5 km s$^{-1}$ and 32 for COS for 15 km s$^{-1}$. When binned to the same velocity interval the S/N improvement in the COS observations compared to the STIS observations is a factor of 2.2 and 1.8 for the O VI λ1031 and H I λ1031 absorption lines, respectively. These S/N measurements were directly determined from the rms errors in the continuum fits.

Simultaneous Voigt profile fits to the COS+ STIS observations of O VI λλ1031, 1037 are displayed in Figure 3. The fit results are listed in Table 2. The fit software utilized the code developed by Fitzpatrick & Spitzer (1999) adapted to use the wavelength dependent COS line spread functions from Kriss et al. (2011) and the two-component line spread function from the STIS Instrument Handbook (Bostroem et al. 2010). Since the wavelength calibration of the STIS observations are more reliable than for the COS observations, the fit code was allowed to vary the velocity of the COS observations as part of the fit process. That insured that the COS observations were brought into the wavelength system of the STIS observations.

Simultaneous Voigt profile fits to the COS+ STIS + FUSE observations to the H I λλ 1215 to 923 series absorption is displayed in Figures 4 and 5 and listed in Table 2. In this case the fit code utilized a 20 km s$^{-1}$ wide line spread function with a Gaussian shape for fitting the FUSE observations. Details regarding the fitting of the H I Lyman series absorption are discussed in section 5.

## 5. PROPERTIES OF THE BLA

The greatly increased S/N for the COS observations of H I λλ1215, 1025, 972 over those available from STIS permits a search for a BLA possibly associated with the O VI and Ne VIII absorption. By employing a three component fit to the H I series observations, Savage et al. (2005) noted the possible presence of a BLA with logN(H I) = 13.31±0.30: , v = 7 km s$^{-1}$, and b(H I) = 97 km s$^{-1}$. The uncertain derived properties were consistent with those expected for the H I in a plasma with logT = 5.73 derived from the O VI and Ne VIII observations. With the new COS observations we have repeated the H I component fit process. In Table 2 we list the results of two and three component Voigt profile fits to the COS, STIS and FUSE observations of the H I Lyman series absorption using the Kriss et al. (2011) LSF. When performing the fits, the COS and FUSE absorption line velocities were allowed to vary while the more reliable STIS velocities were fixed. No constraints were placed on the parameters of the components in each fitting process. The fits were performed over the velocity range from -350 to 350 km s$^{-1}$. The three component free fit results are displayed in Figure 4. Enlarged displays of the 2 and 3 component fits to H I 1215, 1025 are shown in Figure 5. The value of $\chi_\nu^2$ for the full set of lines decreases from 0.89 to 0.87 when adding in a third component. The value of $\chi_\nu^2$ for the COS observation of H I λ1215 line alone is 1.22 for the two component fit and decreases to 0.94 for the three component fit. A comparison of the fits in Figure 5 reveals the third component is required to fit the positive and negative velocity wings of H I λ1216. Therefore the primary evidence for the presence of a BLA is provided by the new COS observations of H I λ1215. The derived velocities of the strongest narrow H I component are -20.6±0.9 and -16.0 ±1.1 km s$^{-1}$ for the two and three component fits, respectively. Both values are consistent with the velocities found for C III, and Si III (see Table 2). However, the derived properties of the second weaker and narrower H I component is strongly affected by including or excluding the third broad component in the fit process. This is because of the strong overlap between the velocity of this component and the velocity of the BLA.

The derived properties of the BLA from the fit are v = 11.2±2.6 km s$^{-1}$, b(H I) = 71.6±5.5 km s$^{-1}$, and logN(H I) = 13.87±0.08. The statistical significance of the detected BLA can be estimated from its logarithmic error of 0.08 dex to be ~ 5 sigma. However, the errors for the BLA listed above are statistical errors and do not allow for uncertainties in the component structure of the absorption or for the possible full range of uncertainties in the continuum placement. If the kinematical properties of the absorber are more complex than the assumption of three simple Voigt profiles, the derived values of the parameters for the BLA will change. However, it is difficult to estimate the impact of our uncertain knowledge of the true kinematic properties of the absorber other than to treat the statistical errors derived for the BLA with caution. The broad wings on the COS observations of the H I λ1215 line at negative and positive velocity could be fitted by adding two narrower H I components rather than a single broad component. However, there is no evidence for additional metal line absorption that might be associated with two outlying H I components. In contrast it is reasonable to expect the existence of a single broad H I absorption line associated with the O VI and Ne VIII absorption. We therefore believe our explanation of the H I λ1215 absorption involving two narrow components tracing cool gas and a single BLA makes the most sense.

By choosing alternate but acceptable continuum fits to the COS H I λ1215 line and repeating the three component fit process we obtained BLA fit parameters within the listed ±1σ errors listed in Table 2 except for the BLA Doppler parameter where the acceptable range is from b = 72±6 to 80±5 km s$^{-1}$.

In the following sections we will adopt the three component fit results for the BLA listed in Table 2 but have adjusted the errors to include the uncertainties in the continuum placement. We therefore adopt v(BLA) = 11±3 km s$^{-1}$, b(BLA) = 72 (+13, -6) km s$^{-1}$, and logN(H I) = 13.87±0.08. Note that the velocity error listed here does not include the COS to STIS calibration error of ±5 km s$^{-1}$.

## 6. THE PHOTOIONIZED GAS

The derived properties of the ions tracing photoionized gas in the absorber at z = 0.20701 observed by COS including H I, C III, N III, and Si III have not changed enough to warrant repeating the detailed photoionization analysis of Savage et al. (2005) which also included FUSE observations of O III, O IV, S VI and a possible detection of S V. The N V detected by COS and not by STIS probably occurs in the photoionized gas. The model of Savage et al. (2005) that best explains the observations of O IV, O III, C III, N III, Si III, S V, and S VI along with the strongest narrow component of H I absorption predictes that logN(O VI) = 13.15 and logN(N V) = 13.2 should exist in the photoionized gas. We observe logN(O VI) = 14.38±0.01 and log N(N V) = 13.24±0.05. The O VI does not originate in the photoionized gas but the N V probably does arise in the photoionized gas. The observed N V absorption has b(N V) = 13.9±3.6 km s$^{-1}$ which is 2.5 times narrower than the O VI absorption with b(O VI) = 34.4±0.7. This line width difference implies the O VI and N V do not occur in the same plasma which is consistent with the photoionized gas origin for N V.

The photoionization calculations of Savage et al. (2005) adopted the extragalactic background (EGB) radiation field from AGNs of Haardt & Madau (1996) which has a mean specific intensity of $j_\nu = 1.9 \times 10^{-23}$ erg cm$^{-2}$ s$^{-1}$ Hz$^{-1}$ sr$^{-1}$ at 912 Å at z = 0.207. The presence of three galaxies near z = 0.20701 (see §7) with L = 0.05, 0.25 and 0.25L$_*$ impact parameters ρ = 37.9$h_{70}^{-1}$, 109 $h_{70}^{-1}$, and 281 $h_{70}^{-1}$ physical kpc, respectively are unlikely to produce enough radiation to significantly modify the mean intensity from the extragalactic background alone. This conclusion follows by scaling the results for the ionizing radiation contributions from the Milky Way for high velocity clouds located different distances from the Milky Way as presented by Fox et al. (2005a) to the galaxy luminosities and absorber/galaxy distances appropriate for the z = 0.20701 system. In all cases the galaxies are assumed to have the spectral energy distribution derived for the Milky Way but scaled to the different luminosities. The assumptions behind the energy distribution appropriate for the Milky Way are discussed in Fox et al. (2005a). We find for the three galaxies that $j_\nu$ (EGB)/$j_\nu$(galaxy) at 912 Å is 25, 40 and 270, respectively, implying a negligible galaxy contribution to the radiation field in all cases

A more important modification to the radiation field adopted by Savage et al. (2005) relates to the assumptions about the relative mixture of galaxies and AGNs when estimating the integrated extragalactic background and the effects of Lyman He II absorption (Madau & Haardt 2009). The radiation field obtained by only including the background from AGNs differs considerably from the background derived from a mixture of star forming galaxies and AGNs and from the inclusion of absorption by He II in the IGM. This uncertainty in the extragalactic radiation background introduces a large systematic uncertainty in the various parameters derived from these simple one zone photoionization models. For example, Howk et al. (2009) note that EGB uncertainties can introduce ~0.3-0.5 dex systematic errors in the derived abundances. Fully exploring the implications of these and other proposed EGB modifications is beyond the scope of this paper which is mostly focused on the properties of the collisionally ionized warm absorber traced by O VI, Ne VIII, and the BLA.

## 7. THE COLLISIONALLY IONIZED GAS

Savage et al. (2005) showed that collisionally ionized gas is required to explain the presence of Ne VIII and O VI in the absorption system at z = 0.20701. Photoionization by the extragalactic background was ruled out because the absorption would need to occur in an extremely low density plasma to make the ionization parameter large enough to produce Ne VIII. The large required path length of 11 Mpc (Savage et al. 2005) would produce a Hubble flow broadening significantly larger than the observed Ne VIII and O VI line widths.

In this section we use the new COS results for the BLA and O VI to constrain the ionization conditions in the collisionally ionized gas. The measurements we are trying to understand are as follows:
i. v(BLA) = 11±3±5 km s$^{-1}$, b(BLA) = 72 (+13,-6) km s$^{-1}$, and logN(H I) = 13.87 ±0.08.
ii. v(O VI) = 0.9±1.4±5 km s$^{-1}$, b(O VI) = 34.4±0.7 km s$^{-1}$, and logN(O VI) = 14.38±0.01.
iii. v(Ne VIII) =-7±6±5 km s$^{-1}$, b(Ne VIII) = 22.6±15.0, and log N(Ne VIII) = 13.89±0.11.

The result for Ne VIII is from Savage et al. (2005).

The O VI line shape is very well described by a Voigt profile with b(O VI) = 34.4±0.7 km s$^{-1}$ (see Fig. 3). If the O VI line broadening is dominated by thermal broadening the implied temperature from $b_T = 0.129 (T/A)^{0.5}$ is log T = 6.06±0.01 where A =16 is the atomic mass number.

If the line broadening in the BLA with b(H I) = 72 (+13,-6) km s$^{-1}$ is mostly from thermal line broadening the implied temperature in the gas is logT = 5.49(+0.15, -0.07) which is close to the temperature where O VI peaks in abundance in collisional ionization equilibrium.

If the O VI and the H I in the BLA exist in the same gas and we ignore the marginal 10±10 km s$^{-1}$ velocity offset between the two species, we can simultaneously solve for the thermal and non-thermal (turbulent) contributions to the broadening assuming $b_O = b_T^2 + b_{NT}^2$ where O, T, and NT refer to the observed, thermal, and non-thermal components to the line broadening, respectively. This equation assumes the non-thermal broadening is well described by a simple Gaussian function. Using $b_O$(O VI) = 34.4±0.7 km s$^{-1}$ and $b_O$(H I) = 72 (+13, -6) km s$^{-1}$ we obtain, logT = 5.41(+0.18, -0.11) and $b_{NT}$ = 30.4 (+1.6, -3.4) km s$^{-1}$. This would imply that most of the broadening of the O VI absorption is non-thermal and that the non-thermal broadening produces a profile very well described by a simple Gaussian optical depth profile. In contrast most of the broadening of the BLA is from thermal motions.

Another approach for estimating the temperature of the gas is to assume the O VI and Ne VIII exist in the same plasma. With logN(Ne VIII) = 13.89±0.11 and logN(O VI) = 14.38±0.01 we obtain log[N(Ne VIII)/N(O VI)] = -0.49±0.11. In a plasma in collisional ionization equilibrium (CIE) with a logarithmic solar abundance ratio of Ne to O of -0.76±0.09 dex from Asplund et al. (2009), the value log[N(Ne VIII)/N(O VI)] = -0.49±0.11 is obtained for log T = 5.68±0.01. At this high a temperature the effects of non-equilibrium cooling are small. For example, if the gas has a solar metallicity the non-equilibrium cooling calculation of Gnat and Sternberg (2007) for isobaric cooling achieves the observed Ne VIII to O VI ratio at log T = 5.67±0.01. However, the elemental abundance estimate below implies the metallicity is closer to -1 dex where there is no difference between CIE and non-equilibruim collisional ionization models.

A potential problem with using the observed Ne VIII to O VI column density ratio to estimate the temperature of the gas is associated with the current uncertainty in the solar ratio of Ne to O and assuming the abundances in the observed plasma are in the solar ratio. However, the ratio of Ne VIII to O VI is sensitive to temperature. Allowing the error on the solar ratio to increase from 0.09 to 0.30 dex only increases the error on log T to 0.02 dex. We therefore adopt log T = 5.68±0.02 dex for the temperature inferred from the observed value of log N(Ne VIII)/ N (O VI).

At log T = 5.68±0.02 and for solar values of [N/O] and [S/O], logN(O VI) = 14.38±0.01 implies the associated column densities of N V and S VI in CIE should be log N(N V) = 12.80±0.01 and logN(S VI) = 12.60±0.01. These are smaller than the observed values of logN( N V) = 13.24±0.05 and logN (S VI) = 12.78 ±0.11 (from Savage et al. 2005). The observed column densities of N V and S VI are consistent with their origin in the cool photoionized absorber.

The analysis discussed above is extremely simplistic. It assumes a single isothermal absorbing layer and that O VI, Ne VIII, and the BLA are uniformly distributed in that layer. It is therefore not too surprising to see that the two methods for evaluating temperatures give somewhat different results. However, both methods point toward gas with logT ~ 5.4 to 5.7. In this temperature range the relative values of O VI and Ne VIII are very sensitive to temperature and therefore provide a more reliable temperature estimate than the method using the different line widths of H I and O VI. More realistic non-isothermal models could easily be constructed to explain the relatively small temperature differences obtained from the two methods. However, with so few constraints it makes little sense to construct more elaborate thermal models.

We can obtain an estimate of the abundance of oxygen in the plasma but the result is sensitive to the adopted temperature of the plasma. The abundance estimate assumes O VI and Ne VIII exists in the H I traced by the BLA. For logT = 5.68 ±0.02 from the observed O VI to Ne VIII column density ratio, along with logN(H I) = 13.87±0.08, logN(O VI) = 14.38±0.01, we obtain [O/H] = -0.89±0.08±0.07 for CIE. The first error is the statistical error from the observed column densities and the second error is from the uncertainty in log T introduced by the assumption that [O/Ne] is given by the solar value.

The total baryonic column density of the plasma is logN(H) = 20.06 ±0.09 for logT = 5.68±0.02. The derived abundances and total baryon content are hardly affected by including the effects of non-equilibrium ionization is a cooling plasma because the temperature of the gas is relatively high and the abundances are relatively low. For log T > 5.5 the non-equilibrium effects are hardly noticeable (see Gnat & Sternberg 2007). Also, with lower abundances, the cooling times become longer and the non-equilibrium effects become less significant (Gnat & Sternberg 2007).

The very large baryonic column density of the collisionally ionized plasma with logN(H) = 20.06±0.09 is ~22 times larger than for the photoionized gas estimated by Savage et al. (2005) to have logN(H) ~ 18.7. The total warm gas column density in the system toward HE026-4110 is somewhat smaller than the value logN(H)= 20.42(+0.13, -0.17) in the hot plasma with logT = 6.07(+0.09, -0.12) at z = 0.22601 found toward HE 0153-4520 (Savage et al. 2011).

## 8. THE ABSORBER GALAXY CONNECTION

Mulchaey & Chen (2009) have identified three galaxies associated with the absorber at z = 0.20701 toward HE 0226-4110. The properties of the galaxies as inferred from their work are listed in Table 3. The galaxy system contains two $0.25L_*$ disk galaxies and a $0.05L_*$ disk galaxy all within 200 km s$^{-1}$ and $\rho < 290h_{70}^{-1}$ physical kpc of the absorber. An R band image of the QSO/galaxy arrangement and the individual galaxy spectra are shown in Figure. 1 of Mulchaey & Chen (2009). Galaxy A and B lie closest to the QSO line of sight with impact parameters of $37.9h_{70}^{-1}$ and $109h_{70}^{-1}$ kpc, respectively. A and B are both disk galaxies with emission lines indicating normal star formation. Galaxy C is a disk galaxy with an absorption line spectrum tracing an evolved stellar population. It has an impact parameter of $281h_{70}^{-1}$ kpc.

The Mulchaey & Chen (2009) redshift survey was relatively deep. They can therefore rule out the existence of additional galaxies more luminous than $0.02L_*$ with $\rho < 407h_{70}^{-1}$ kpc. Mulchaey & Chen (2009) argue that A and B probably share a common dark matter halo and that the absorption system probably occurs in the circumgalactic environment of A and B.

## 9. ORIGINS OF THE ABSORPTION SYSTEM

We discuss possible sites of absorption for the z = 0.20701 system toward HE 0226-4110. The low density circumgalactic regions around galaxies are likely to be complex. With a single line of sight through such a region it is not easy to clearly discriminate among the many possible sites of origin of an absorber even when good knowledge exists regarding the locations and properties of the galaxies near the redshift of the absorber. In the following we discuss the strong and weak aspects of several different but possible origins for the absorber. Our discussion expands on the absorber origins remarks of Mulchaey & Chen (2009).

### 9.1 An Extended Hot Halo Origin

Mulchaey & Chen (2009) estimate that Galaxy B and A have virial radii of $196h_{70}^{-1}$ and $133h_{70}^{-1}$ kpc, respectively, and Virial temperatures of $4 \times 10^5$ and $2 \times 10^5$ K, respectively. When trying to decide if the absorber is tracing possible warm extended halo(s) of one or both galaxies, it is important to estimate the expected line widths for various metal line absorbers possibly associated with warm gaseous halos. The expected line widths will depend on the origins of the heating of the warm gas and when in its evolution the warm gas is observed.

A key issue is the expected amount of turbulent versus thermal Doppler broadening of the absorption produced by Ne VIII, O VI and H I in warm extended halos. If the warm halo is formed as the result of the assembly of the galaxy, the kinetic energy of the virial motions will be converted into thermal energy by the shock heating of the gas. Once the gas is heated to a temperature, T, the particles in the gas will assume a line of sight velocity distribution from thermal motions characterized by the thermal Doppler parameter $b_T = 0.129 (T/A)^{0.5}$ km s$^{-1}$. For T = $3 \times 10^5$ K, H I and O VI will have thermal Doppler parameters of 70.7, and 17.7, km s$^{-1}$, respectively. We will restrict our discussion here to H I and O VI because the observed parameters for O VI are of much higher quality than those for Ne VIII. The observed Doppler parameters will be affected by the turbulence and the kinematic flows or differential rotation in the absorbing structure. For the case above, if the thermal energy content in the warm halo is the same as the turbulent energy content then $b_{NT}$ for all the ions will be the same as $b_T$ (H I) = $b_T(H^+)$ = 70.7 km s$^{-1}$ and the observed Doppler parameters for the H I and O VI would be 100 and 72.9 km s$^{-1}$ respectively, assuming the thermal and turbulent broadening parameters add in quadrature. However, we observe b(H I) = 72 (+13, -6) and b(O VI) = 34.4±0.7 km s$^{-1}$ which implies log T = 5.41(+0.18, -0.11), $b_T$ (H I) = 65.4 (+15.1, -7.8) km s$^{-1}$, $b_T$ (O VI) = 16.3 (+3.8, -1.9) km s$^{-1}$, and $b_{NT}$ = 30.4 (+1.6, -3.4) km s$^{-1}$. In the gaseous medium observed, the energy content of the turbulence is 4.6 times smaller than the thermal energy content of the gas. A level of turbulence corresponding to $b_{NT}$ = 30.4 km s$^{-1}$ seems small when looking through a shock heated halo along a ~200 kpc path length. Any other sources of motion such as kinematic flows or rotation of the structure will make the inferred level of turbulence even smaller.

With an oxygen abundance [O/H] = -0.89±0.08±0.07 from the COS measurements, the gas in the absorber is somewhat enriched and could be consistent with a warm halo interpretation.

If the warm halo explanation is correct it would be necessary to also explain the 196 km s$^{-1}$ velocity difference between the absorber and galaxy B. This offset could occur from rotation of the halo. However, galaxy B appears to be roughly face on. The velocity problem is less severe if the absorber traces the halo of galaxy A for which the velocity difference between the absorber and the galaxy is 127 km s$^{-1}$.

For the warm halo origin for the O VI, Ne VIII and BLA to be correct it is also necessary to explain the origin of the cooler photoionized gas which appears to be closely kinematically connected to the hotter gas absorber because its velocity difference is small. This cooler absorber could be a condensation in the warm halo or trace denser gas associated with galaxy A or B. However, in these cases it seems unlikely that the dense absorber would have nearly the same velocity as the warm gas absorber if the warm gas absorption occurs over several hundred kpc. From the above discussion we believe the z = 0.20701 system is tracing a physically smaller structure in the vicinity of Galaxy A or B rather than a highly extended warm halo. Mulchaey & Chen (2009) arrived at a similar conclusion basing their arguments on the very uncertain b-value of the Ne VIII absorption.

### 9.2 Cool/Hot Gas Interface Origin

Turbulent and conductive interfaces between gaseous regions with log T ~ 4.3 and log T > 6 are places where the highly ionized ions tracing gas with log T ~ 5.3 -5.7 can arise (Cowie & McKee 1977; Mckee & Cowie 1977; Boehringer & Hartquist 1987; Begelman & Fabian 1990; Borkowski et al. 1990; Slavin et al. 1993; Gnat et al. 2010; Kwak & Shelton 2010; Kwak et al. 2011). The strong kinematic connection between O VI and such cooler gas tracers as O I and C II (Cowie et al. 1979; Savage & Lehner 2006; Bowen et al. 2008; Lehner et al. 2011) observed in the disk and local interstellar medium of the Milky Way provides support to the interface explanation for the origin of O VI in these galactic sites. However, a single conductive interface only produces log N(O VI) ~ 12.5 (Borkowsky et al. 1990; Gnat et al 2010). Therefore, ~75 interfaces would be required to produce the total observed O VI column density in the z = 0.20701 absorber. The interactions produced by a cool cloud moving with a speed of 100 to 200 km s$^{-1}$ through a hot halo might produce enough turbulence to produce the large number of interfaces when observing through the entire structure. However, the process would need to produce a highly symmetrical Gaussian broadened O VI absorption line with b(O VI) = 34.4±0.7 km s$^{-1}$. Young evaporating conductive interfaces should have the same elemental abundances as the cooler matter being evaporated. Old condensing interfaces will have the elemental abundances of the hot gas being condensed. In the case of the absorber at z = 0.20701, the cool photoionized gas has [Z/H] ~ -0.5±0.2 while the gas traced by O VI, Ne VIII and the BLA has [O/H] = -0.89±0.08±0.07. The marginally different abundances imply the hotter gas may not arise from the conductive heating of the cooler gas but instead from the cooling of the hot (unseen) exterior medium in the interfaces between the cool gas and the hot gas.

The presence of galaxy A close to galaxy B raises the possibility that galaxy A may be producing a stream of gaseous matter equivalent to the Magellanic Stream around the Milky Way. Strong O VI absorption near the velocity of cooler gas has been detected in high velocity clouds (HVCs) including the Magellanic Stream (Sembach et al. 2003; Fox et al. 2010). A viable interpretation for the origin of the O VI in HVCs is that the O VI occurs in the turbulent interfaces between the cool gas of the cloud and the hot gas of an extended corona around the Milky Way (Fox et al. 2004; Collins et al. 2007; Kwak et al. 2011) . With logN(H I) ~15.3 in the cool gas, the absorber at z = 0.20701 resembles the Galactic highly ionized high velocity clouds where the column density of H I is too small to be seen in 21 cm emission. In the study of highly ionized HVCs in the Milky Way by Fox et al. (2005b) it was found that  <b(O VI)> = 38±10(STDEV) and <logN(O VI)> = 13.83±0.36 (STDEV).   In the z =0.20701 absorber b(O VI) =  34.4±0.7 km s$^{-1}$  and logN(O VI) = 14.38±0.01.  Although the O VI column density is ~3 times larger than in the average highly ionized HVC,  the line width is very similar.  A turbulent cool/hot gas interface origin of the z = 0.20701 absorber toward HE 0226-4110 is therefore a plausible interpretation.  This is the origin favored  by Muchaey & Chen (2009).

### 9.3 Other Possible Origins.

Mulchaey  & Chen (2009) provided arguments that the absorption system is probably not due to an starburst driven outflow from galaxy A.  The restframe Hα equivalent width of ~13 Å implies the galaxy is not a starburst galaxy.  In addition  the major axis of the galaxy is along the direction to the QSO  while an outflowing wind would be expected to be along the minor axis.  The relatively low metallicity of the warm gas  also casts doubt on an origin in a starburst driven outflow.

The very close proximity of galaxies A and B to the QSO line of sight makes it unlikely that the absorption arises in regions unrelated to the two galaxies.   Although the circumgalactic environment of the two galaxies could connect to an intergalactic filament containing warm gas, the small difference in velocity between the warm and cool gas absorpion suggests the filament would also need to contain cool photoionized  gas with a relatively high metallicity   ([X/H] ~ -0.5) to explain the absorption system.  An origin in the warm gas of a group of galaxies possibly associated with the three galaxies would also need to explain the small velocity offset between the warm and cool gas.  In addition,  in both of these situations it is  difficult to understand how absorption extending over large  filament scale or group scale  path lengths could produce a relatively symmetric and narrow O VI absorption line with b = 34.4±0.7 km s$^{-1}$. Finally, the recent absorber/galaxy connection study at cz < 2500 km s$^{-1}$ by Wakker & Savage (2009)  that is sensitive to galaxies with L > 0.1L$_*$ finds that each L > 0.1L$_*$ field galaxy and about half of L>0.1L$_*$ group galaxies have an O VI absorber within an impact parameter of 350 kpc.  Most O VI absorbers therefore appear to be associated with the circumgalactic environments of galaxies with L>0.1L$_*$. Our knowledge of the absorber/galaxy connection should improve rapidly over the next few years as the results from HST/COS programs specifically designed to explore this connection are published.

### 10.  SUMMARY

S/N~ 20 to 40  observations  of the QSO HE 0226-4110  (z$_{em}$ = 0.495) with COS from 1134 to 1796 Å with a resolution of ~17 km s$^{-1}$ are used to study the multi-phase absorption system at z = 0.20701 containing O VI and Ne VIII.  The system was previously studied by Savage et al. (2005) with lower S/N observations with FUSE and STIS.   The COS observations provide more reliable measures of the H I and metal lines present in the system from 1134 to 1796 Å.  The clear detection of a BLA associated with the O VI and Ne VIII allows a determination of the ionization conditions, the elemental abundances and the total hydrogen column density in the warm ionized gas traced by the absorber.

   1.  The COS observations provide significantly higher S/N measurements of  H I λλ1215 to 949,  O VI λλ1031, 1037, N V  λλ1238, 1242,  C III λ977, N III λ990, and Si III  λ1206 than for the earlier STIS observations of Savage et al. (2005).

   2. Multi-component profile fits to the COS, STIS and FUSE observations of  H I λλ1215 to 923 reveals the presence of a BLA  with  v = 11±3 km s$^{-1}$, b = 72 (+13,-6) km s$^{-1}$, and logN(H I) = 13.87±0.08.  The BLA is likely recording the trace amount of H I that exists in the collisionally ionized gas that is responsible for the O VI and Ne VIII absorption with log N(O VI) = 14.38±0.01 and log N(Ne VIII) = 13.89±0.11.  The BLA was not detected in the lower S/N STIS observation of Lyman α.

   3. The new COS measurements support the conclusions of Savage et al. (2005) regarding the properties of the cooler photoionized gas in the system with T ~2.1x10$^4$ K, $n_H$ ~ 2.6x10$^{-5}$ cm$^{-3}$, P/k ~ 0.5 cm$^{-3}$, logN(H) ~  18.7, N(H I)/N(H) ~ 2.5x10$^{-4}$, and [Z/H] = -0.5±0.2.   The COS detection of N V with log N(N V) = 13.22±0.05  is consistent with the expected column of  N V in the photoionized gas.  Unfortunately there are 0.3 dex systematic errors in the  derived abundances because of uncertainties in the photoionizing extragalactic radiation field.

   4.   The absorption from O VI, Ne VIII, and the BLA points to the existence of collisionally ionized gas with log T ~ 5.7.  The total hydrogen column in the collisionally ionized gas, logN(H) = 20.06±0.09, exceeds that in the photoionized gas by a factor of ~22. The oxygen abundance in the collisionally ionized gas  is  [O/H] = -0.89±0.08±0.07.

   5. Mulchaey & Chen (2009)  have identified three galaxies associated with the absorber at z = 0.20701 toward HE 0226-4110.  The galaxy system contains two 0.25L$_*$ disk galaxies and a  0.05L$_*$ galaxy all within 200  km s$^{-1}$ and  ρ < 200h$_{70}^{-1}$ physical kpc of the absorber.   We discuss the evidence for and against the identification of the possible physical sites of origin of the photoionized and collisionally ionized gas.  While the absorption likely occurs in the circumgalactic (halo) environment of the three galaxies it is difficult to establish with certainty the actual site and extent of the absorption.  However, as suggested by Mulchaey & Chen (2009) it is plausible that the cool photoionized gas in the absorber traces a stream of material moving through a hot medium and the O VI, Ne VIII, and BLA absorption occurs in a large number of turbulent interfaces that would be formed in the interaction region.  A similar process has been proposed to explain O VI absorption in the Magellanic Stream and the Galactic highly ionized HVCs.

Acknowledgements:  We thank the many people involved with building COS and determining its performance characteristics.   A number of helpful comments provided by the referee allowed us to improve our manuscript. BDS and AN acknowledge funding support from NASA through the COS GTO contract to the University of Wisconsin-Madison through the University of Colorado.  This work was supported by NASA grants NNX08AC146 and NAS5-98043 to the University of Colorado at Boulder

Table 1. COS G130M and G160M Integrations of HE0226-4110[a]

| MAST ID | Date[a] | Grating | $\lambda_c$ [Å] | $\lambda_{mim}$ [Å] | $\lambda_{max}$ [Å] | t [sec] |
|---|---|---|---|---|---|---|
| LB6803040 | 2/03/10 | G130M | 1291 | 1137 | 1430 | 1595 |
| LB6803050 | 2/03/10 | G130M | 1300 | 1146 | 1440 | 1452 |
| LB6803060 | 2/03/10 | G130M | 1309 | 1156 | 1449 | 1452 |
| LB6803070 | 2/03/10 | G130M | 1318 | 1165 | 1458 | 1452 |
| LB6803080 | 2/03/10 | G130M | 1291 | 1137 | 1430 | 824 |
| LB6803090 | 2/03/10 | G160M | 1577 | 1389 | 1749 | 2040 |
| LB68030A0 | 2/03/10 | G160M | 1589 | 1401 | 1761 | 1420 |
| LB68030B0 | 2/03/10 | G160M | 1600 | 1412 | 1772 | 1420 |
| LB68030C0 | 2/03/10 | G160M | 1611 | 1424 | 1784 | 1457 |
| LB68030D0 | 2/03/10 | G160M | 1623 | 1436 | 1796 | 1457 |

[a]The table lists the MAST exposure ID number, the date of observation, the COS grating, the central set-up wavelength, the minimum and maximum wavelengths covered by the integration, and the exposure time.

Table 2. Profile Fit Results for the z = 0.20701 Absorber[a]

| ion | $\lambda_r$ | v (km s$^{-1}$) | b (km s$^{-1}$) | log N (dex) | [-v, +v] (km s$^{-1}$) | $\chi_\nu^2$ | Note |
|---|---|---|---|---|---|---|---|
| H I | 1215 to 923 | -20.6±0.9±5 | 17.2±0.7 | 15.05±0.02 | [-350, 350] | 0.89 | 1 |
| " | " | 5.4±0.9±5 | 36.6±0.3 | 14.93±0.02 | " | " | 1 |
| H I | 1215 to 923 | -16.0±1.1±5 | 20.6±0.6 | 15.20±0.01 | [-350, 350] | 0.87 | 2 |
| " | " | 27.7±2.0±5 | 19.9±1.5 | 14.52±0.04 | " | " | 2 |
| " | " | 11.2±2.6±5 | 71.6±5.5 | 13.87±0.08 | " | " | 2 |
| O VI | 1031, 1037 | 0.9±1.3±5 | 34.4±0.7 | 14.38±0.01 | [-140, 190] | 0.99 | 3 |
| N V | 1238, 1242 | -4.0±2.0±5 | 13.9±3.6 | 13.24±0.05 | [-140, 190] | 0.81 | 4 |
| C III | 977 | -18.8±2.8±5 | 16.6±1.8 | 13.81±0.05 | [-150,150] | 1.00 | 5 |
| " | " | 11.4±7.3±5 | 19.0±5.4 | 13.28±0.19 | [-150, 150] | " | 5 |
| N III | 990 | -11.5±9.4±5 | 22.7±5.1 | 13.47±0.06 | [-90, 150] | 1.00 | 6 |
| Si III | 1206 | -17.9±3.9±5 | 21.7±2.3 | 12.37±0.03 | [-150, 90] | 1.09 | |

[a]The Voigt profile fit code of Fitzpatrick & Spitzer (1994) and the wavelength dependent COS LSFs from Kriss et al. (2011) were used to obtain the component fit results listed in this table. In all cases the new COS observations displayed in Figure 2 were simultaneously fitted with the STIS and FUSE observations from Savage et al. (2005) using the appropriate line spread functions for each of these instruments. The COS and FUSE velocities were allowed to vary in the fit process in order to make them agree with the more accurate velocities from the STIS observations. The first listed velocity error is the fit statistical error. The second error is the approximate COS to STIS velocity calibration error. Absorption line wavelengths and f-values were taken from Morton (2003). The profile fits for O VI are shown in Figure 3. The component fits to the H I absorption are shown in Figures 4 and 5. Various non-detections based on the COS observations are not listed since they do not provide useful additional constraints on the modeling of the photoionized or collisionally ionized absorbers. The non-detections include C II λλ1036, 1335, N II λ1084, Si IV λ1403 (Si IV λ1393 is blended with H I λ1215 at z = 0.38427), Si II λλ1190, 1193, 1260, Fe III λ1123, and Fe II λ1145. The 3σ equivalent width detection limits are typically 10 to 20 mÅ.

Notes: (1) Two component free fit to H I λλ1215 to 923. The velocities of the COS and FUSE observations were allowed to vary to best fit the more accurate velocities from the STIS observations. The two component fit for the COS and STIS observations of H I λ1215 yields $\chi_\nu^2$ = 1.22 for the individual line. (2) Three component free fit to H I λλ1215 to 923. The fit to the H I λ1215 line for the COS and STIS observations is improved to $\chi_\nu^2$ = 0.94 for the individual line. (3) Single component fit to the COS and STIS observations of O VI λλ1031.1037 absorption. The O VI absorption is well described by a single component Voigt profile. (4). The STIS observation of N V λ1238 was contaiminated by a strong broad feature identified as H I λ1215 at z = 0.23010 by Lehner et al (2006). That feature is not seen in the COS observations. The N V results listed are from COS measurements alone using the COS to STIS wavelength alignment determined from the strong lines near 1483.2 and 1507.0 Å. (5) A two component fit to the COS and STIS observations of C III. The velocity of the second weaker component at v = 10.4±7.3 km s$^{-1}$ is not strongly constrained. (6) N III velocity is uncertain because the COS velocity adjustment was to a very noisy STIS observation.

Table 3. Galaxies associated with the absorber at z = 0.20701 from Mulchaey & Chen(2009)[a]

| Galaxy | z | $\Delta v$ (km s$^{-1}$) | $\rho$ (kpc) | $M_R - 5\log h_{70}$ | $L/L_*$ | $R_H$ (kpc) | $T_{VIR}$ (K) | $M/M_O$ | $M_*/M_O$ | note |
|---|---|---|---|---|---|---|---|---|---|---|
| A | 0.2065 | 127 | $37.9 h_{70}^{-1}$ | -16.43 | 0.05 | $133 h_{70}^{-1}$ | $2 \times 10^5$ | $10^{11}$ | $1.6 \times 10^9 h_{70}^{-2}$ | 1 |
| B | 0.2078 | 196 | $109 h_{70}^{-1}$ | -18.13 | 0.25 | $196 h_{70}^{-1}$ | $4 \times 10^5$ | $10^{11.5}$ | $8.2 \times 10^9 h_{70}^{-2}$ | 1 |
| C | 0.2077 | 171 | $281 h_{70}^{-1}$ | -18.13 | 0.25 | $196 h_{70}^{-1}$ | $4 \times 10^5$ | $10^{11.5}$ | $8.2 \times 10^9 h_{70}^{-2}$ | 2 |

[a] The quantities listed include the galaxy ID, redshift, absorber/galaxy velocity difference, QSO-galaxy impact parameter, absolute R magnitude, $L/L_*$, the dark matter halo radius, the halo virial temperature, the total estimated galaxy mass, the estimated stellar mass, and notes.

Notes. (1) A and B are disk galaxies with emission line spectra indicating normal star formation. (2) C is a disk galaxy with an absorption line spectrum tracing an evolved stellar population.

FIGURES

FIG.1. COS and STIS observations of flux versus heliocentric wavelength from 1235 to 1255 Å for the QSO HE 0226-4110. Absorption features of H I 1025, O VI 1031 and 1037 in the system at z = 0.20701 are identified. The identities of the other ISM and IGM lines are found in Lehner et al. (2006).

FIG. 2. Continuum normalized absorption line profiles versus rest frame heliocentric velocity for selected species found in the COS and STIS observations of the z = 0.20701 system in the spectrum of HE 0226-4110. The COS observations are displayed with a sampling of ~7.1 km s$^{-1}$. The STIS observations are displayed with a sampling of ~3.3 km s$^{-1}$. The species plotted are noted on each panel. Contaminating lines from other absorption systems including ISM lines are displayed with the light lines in the panels for N III $\lambda$989 and Si III $\lambda$1206. The Ne VIII absorption lines in the system are shown in Figures 2, 3, and 4 of Savage et al. (2005)

FIG. 3. Simultaneous Voigt single component profile fits to the O VI $\lambda\lambda$1031,1037 observations from COS and from STIS. The fit results are listed in Table 2.

FIG. 4. Simultaneous Voigt profile fits of three absorption components to the COS, STIS, and FUSE observations of H I $\lambda\lambda$1215 to 923 are displayed. Each panel identifies the Lyman series line and the instrument. In the case of FUSE, the independent observations with the LIF2A and LiF1B channels are fitted separately. Fit lines are displayed for the individual components and for the sum of the three components. The fit results are listed in Table 2.

FIG. 5. An enlarged view of the three (left panels) and two (right panels) component fits to the COS and STIS observations of H I $\lambda\lambda$1215, 1025. Fit lines are displayed for the individual components and for the sum of the three or two components. The residuals (fit –observations) are shown with the light grey lines for the COS H I $\lambda\lambda$1215, 1025 observations.

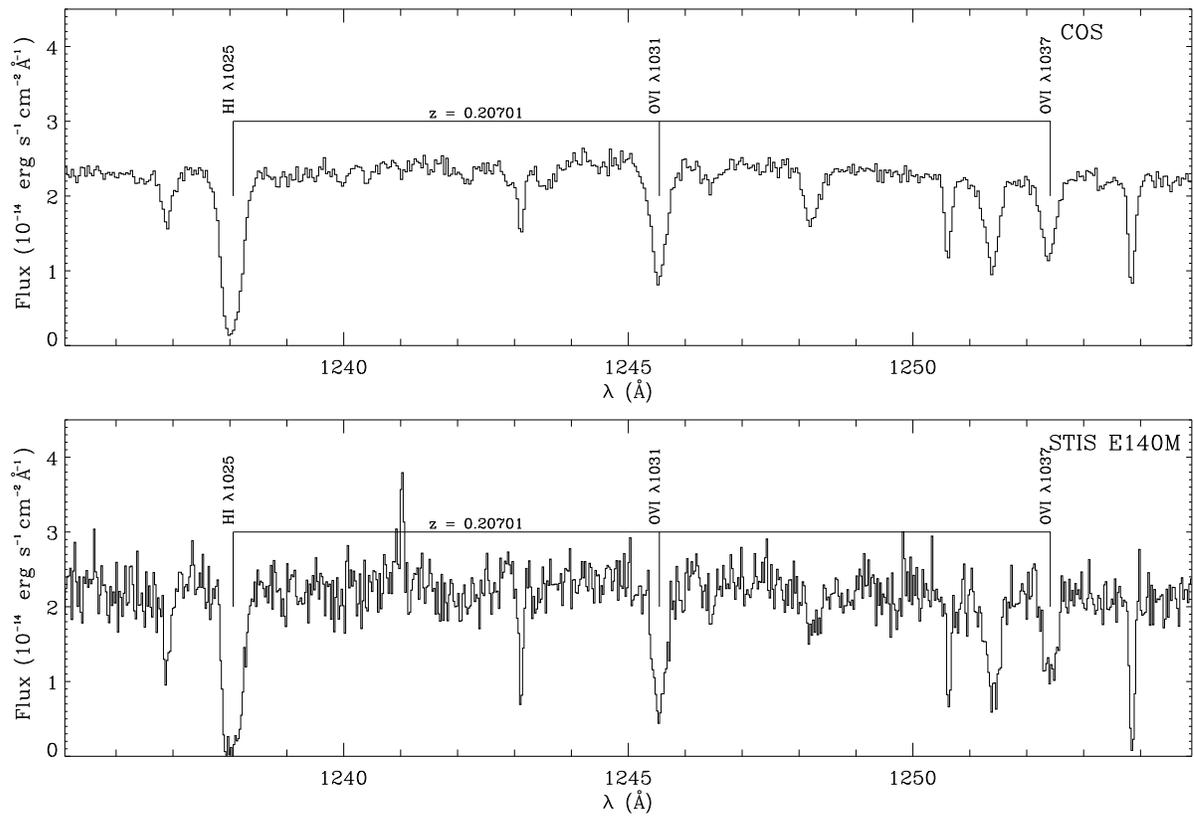

FIG. 1

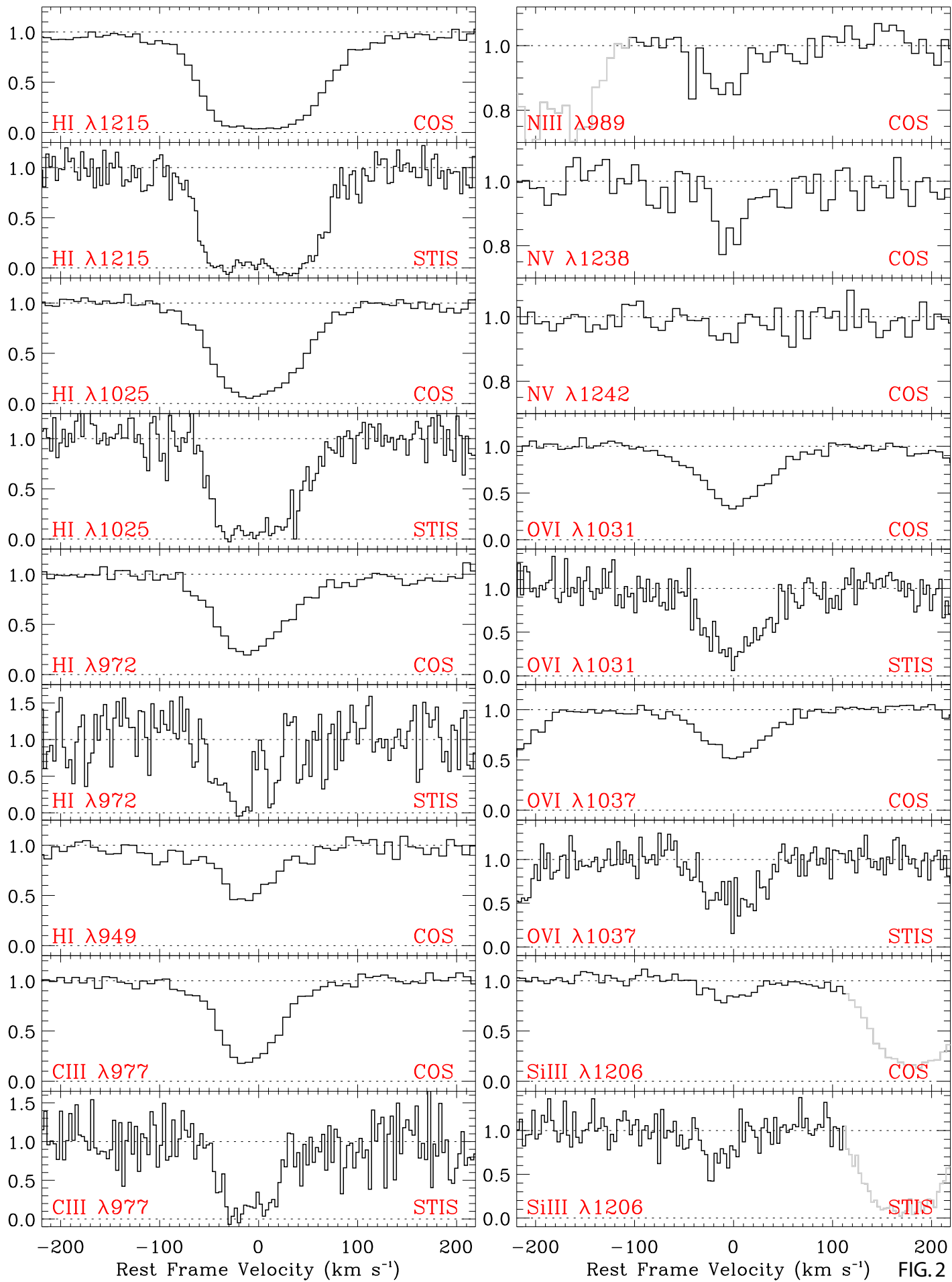

FIG. 2

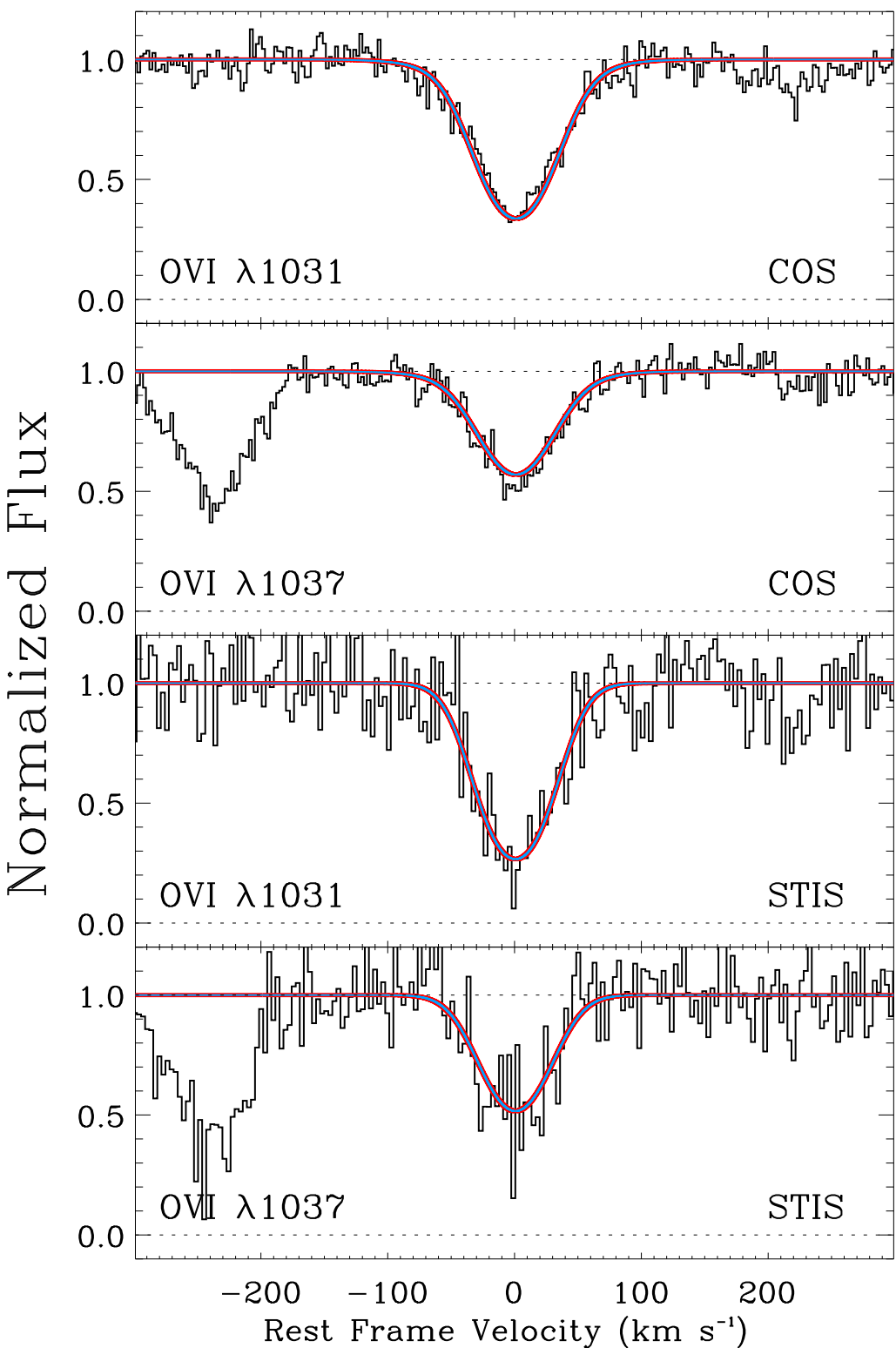
FIG.3

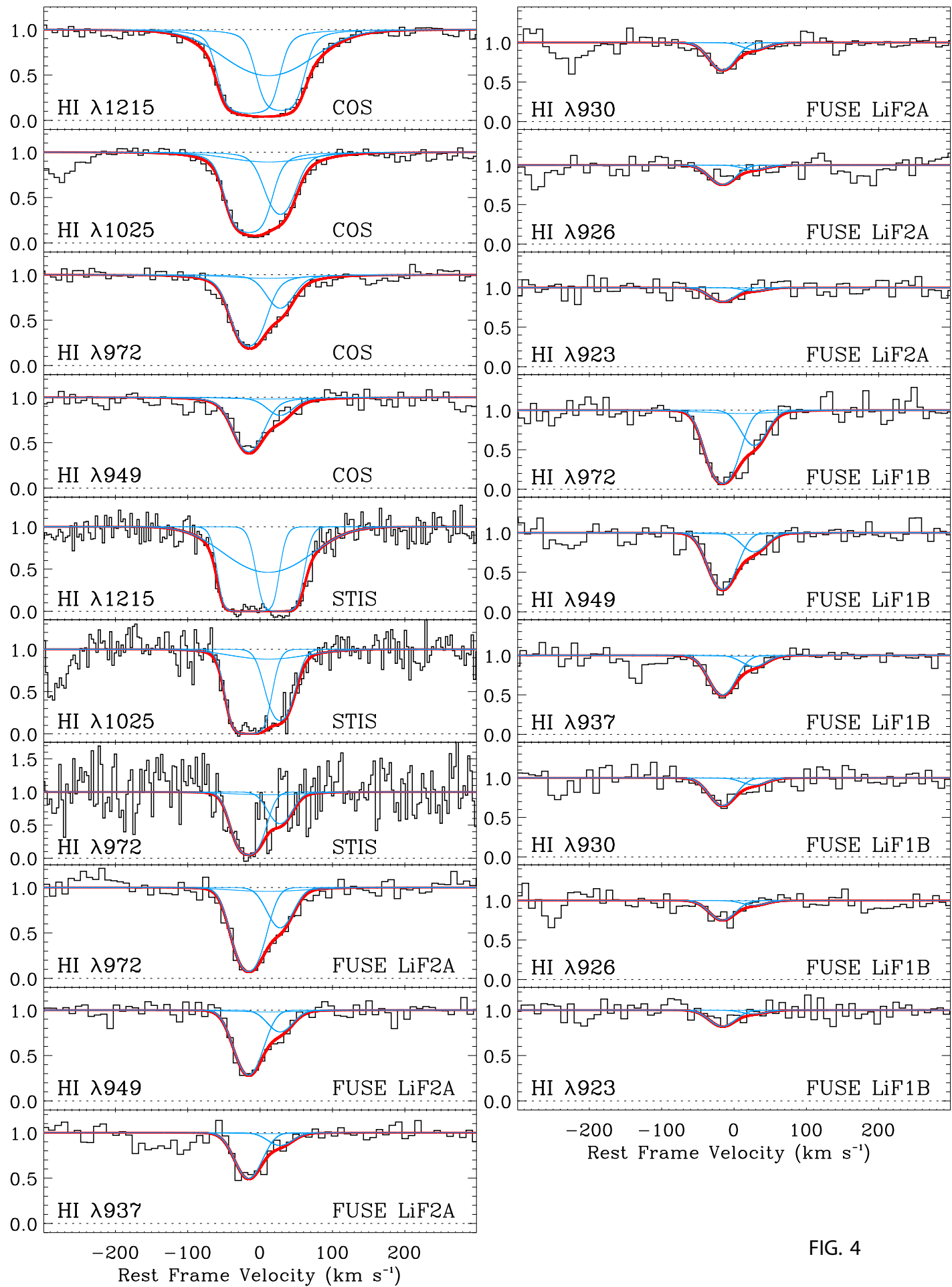

FIG. 4

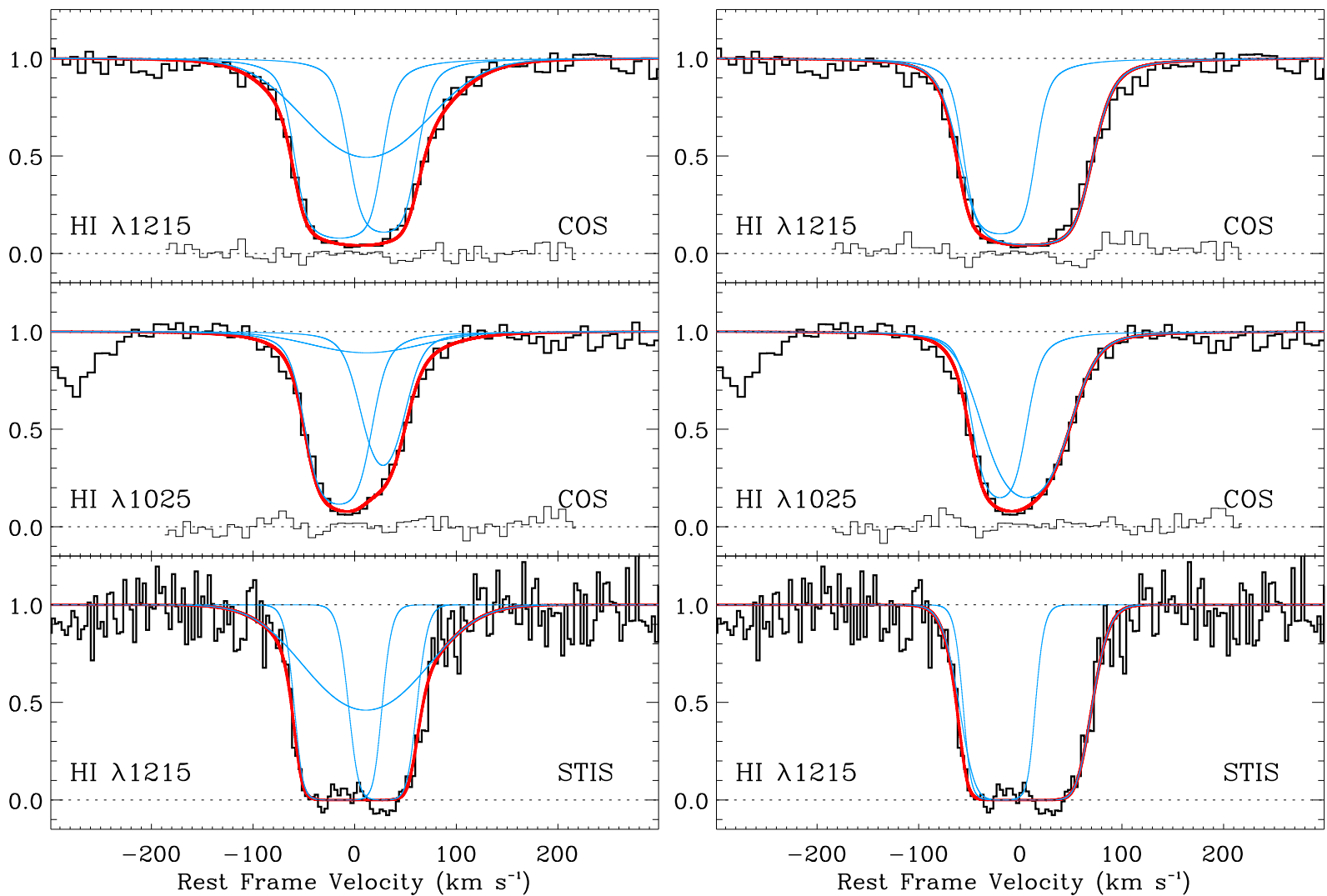

FIG. 5